\documentclass[final,1p,times]{elsarticle}

\usepackage{graphicx}
\usepackage{epsfig}
\usepackage{figsize}
\usepackage{hyperref}
\usepackage{amsmath}
\usepackage{amssymb}
\biboptions{compress}

\journal{Physica B}

\begin{document}

\begin{frontmatter}

\title{Quantum spin Hall insulator on the honeycomb lattice induced by ferromagnetic exchange interaction}

\author[addr1]{Ye-Un An}
\author[addr1]{Song-Jin O\corref{mycorrespondingauthor}}
\cortext[mycorrespondingauthor]{Corresponding author}
\ead{sj.o@ryongnamsan.edu.kp}
\author[addr1]{Kwang-Il Ryom}
\author[addr1]{Il-Gwang Son}
\address[addr1]{Faculty of Physics, Kim Il Sung University, Ryongnam-Dong, Taesong District, Pyongyang,\\
Democratic People's Republic of Korea}

\begin{abstract}
We study the many-body instabilities of correlated electrons on the half-filled honeycomb lattice with enhanced exchange coupling. The system is described by an extended Hubbard model including the next-nearest-neighbor Coulomb repulsion ($V_2$) and the nearest-neighbor exchange interaction ($J$). We use the truncated unity functional renormalization group approach to determine a schematic ground-state phase diagram with higher resolution in the parameter space of $V_2$ and $J$. In the absence of the on-site repulsion and presence of sizable next-nearest-neighbor repulsion and enhanced nearest-neighbor exchange interaction, we encounter the quantum spin Hall phase, the spin-Kekul\'{e} phase, and the three-sublattice and the incommensurate charge-density-wave phases. We propose a scheme for combining consistently the truncated unity functional renormalization group and the mean-field approximation, which is distinct from the conventional one that directly uses the renormalization-group results as an input for the mean-field calculation. This scheme is used to study in detail the quantum spin Hall phase, presenting some characteristics like the bulk gap, the Chern number and the helical edge states.
\end{abstract}

\begin{keyword}
\texttt Topological Mott insulator\sep Functional renormalization group \sep Honeycomb lattice \sep Exchange interaction
\end{keyword}

\end{frontmatter}

\section{Introduction}

The theoretical prediction\cite{ref01, ref02, ref03} and experimental observation\cite{ref04} of the quantum spin Hall insulator (two-dimensional topological insulator) have triggered an intense research on the topological nature of the materials leading to the discovery of novel topological phases of matter and the introduction of new topological invariants\cite{ref05, ref06}. Topological insulators have common features, namely, the nontrivial topological invariants of the bulk bands (e.g., the Chern number and the $Z_2$ invariant) and the occurrence of gapless edge states associated with the invariants by the so-called bulk-edge correspondence\cite{ref07}.

The quantum spin Hall (QSH) state can appear in the systems of independent electrons with sizable spin-orbit coupling, in the absence of external magnetic field. On the other hand, it can also emerge purely from electron-electron interactions. It has been suggested that the quantum anomalous Hall (QAH) and the QSH states could emerge for spinless and spinful electrons, respectively, on the half-filled honeycomb lattice, from a strong next-nearest-neighbor repulsion\cite{ref08}. This scenario has been supported by the early studies\cite{ref09, ref10, ref11}, but many later works\cite{ref12, ref13, ref14, ref15, ref16, ref17, ref18, ref19} excluded it, demonstrating the suppression of the QAH or QSH by conventional charge ordered phases.

Recently, it has been argued that the QSH phase could be induced by a combination of the ferromagnetic exchange and pair hopping interactions\cite{ref20}. There, the QSH appears only when including the appropriate strength of the next-nearest-neighbor repulsion and the direct exchange interaction (the latter gives the equal strength of the ferromagnetic exchange coupling and pair hopping). In other words, pure density-density interactions, or the inclusion of only one of ferromagnetic exchange and pair hopping, cannot cause the QSH effect. Moreover, very recently, having relevance to the QAH state discovered at three-quarter filling of the twisted bilayer graphene\cite{ref21,ref22}, the existence of the topological Mott insulator\cite{ref08} becomes hot issue attracting great interest\cite{ref23}. In the context of these, we think it would be meaningful to determine the phase boundary of the QSH with refined resolution and present more detailed description of this intriguing phase.

As an extension of our previous work\cite{ref20}, in this paper we revisit the half-filled honeycomb lattice with enhanced exchange coupling. We use the truncated unity functional renormalization group (TUFRG) approach\cite{ref24} to study the quantum many-body instabilities of correlated electrons on the system. Focusing on the QSH phase, we consider the extended Hubbard model including the next-nearest-neighbor repulsion $V_2$ and the nearest-neighbor exchange interaction $J$. The TUFRG results are summarized by a schematic phase diagram which has a resolution higher than that in Ref. \cite{ref20}. Another purpose of this paper is to present a detailed description and analysis of the QSH phase detected in the TUFRG calculation, which needs a mean-field (MF) approximation. To this end, we propose a scheme for linking the TUFRG with MF theory, which will be discussed later and employed to calculate the bulk gap, Berry curvature, Chern number, and edge states of the QSH phase.

This paper is organized as follows. In Sec.~\ref{sec2} we give the model Hamiltonian and a brief description of the TUFRG approach, and present the calculated ground-state phase diagram. In Sec.~\ref{sec3} we describe a novel approach for a combination of the TUFRG and the MF theory, intended for the spin channel. Sec.~\ref{sec4} is devoted to an analysis of the QSH phase in the phase diagram, providing with some results of its topological properties. Finally, in Sec.~\ref{sec5} we draw our conclusions.

\section{Model, method and phase diagram} \label{sec2}

Following Ref. \cite{ref20}, we consider an extended Hubbard model of interacting spin-$1/2$ electrons described by the Hamiltonian
\begin{equation}\label{eq01}
H = H_0  + H_{{\rm int}},
\end{equation}
where $H_0$ is a single-particle part with the nearest-neighbor hopping $t$,
\begin{equation}\label{eq02}
H_0  =  - t\sum\limits_{\left\langle {iA,jB} \right\rangle ,\sigma } {(c_{iA\sigma }^\dag  c_{jB\sigma }  + {\rm{H}}{\rm{.c}}{\rm{.}})},
\end{equation}
and $H_{\rm int}$ is an interaction part with the next-nearest-neighbor Coulomb repulsion $V_2$ and the nearest-neighbor exchange interaction $J$,
\begin{equation}\label{eq03}
\begin{split}	
H_{{\rm int} }  = & V_2 \sum\limits_{\left\langle {\left\langle {io,jo} \right\rangle } \right\rangle ,o} {\sum\limits_{\sigma ,\sigma '} {n_{io\sigma } n_{jo\sigma '} } }  + J\sum\limits_{\left\langle {iA,jB} \right\rangle } {\sum\limits_{\sigma ,\sigma '} {c_{iA\sigma }^\dag  c_{jB\sigma '}^\dag  c_{iA\sigma '} c_{jB\sigma } } } \\
 &+ J\sum\limits_{\left\langle {iA,jB} \right\rangle } {\sum\limits_{\sigma ,\sigma '} {(c_{iA \uparrow }^\dag  c_{iA \downarrow }^\dag  c_{jB \downarrow } c_{jB \uparrow }  + {\rm{H}}{\rm{.c}}{\rm{.}})} }.
\end{split}	
\end{equation}
The operator $c_{io\sigma }^\dag  (c_{io\sigma } )$ creates (annihilates) an electron with spin polarity $\sigma$ at lattice site of the sublattice $o$ in the unit cell $i$, and $n_{io\sigma }  = c_{io\sigma }^\dag  c_{io\sigma }$ is the local density operator for the electrons with spin $\sigma$.

In the interaction part of the Hamiltonian, the second and third terms represent the nearest-neighbor ferromagnetic exchange coupling and pair hopping interaction, respectively, while the sums $\sum\nolimits_{\left\langle {iA,jB} \right\rangle }$ and $\sum\nolimits_{\left\langle {\left\langle {io,jo} \right\rangle } \right\rangle }
$ go over nearest and next-nearest neighbors. In view of our objective to search for the QSH state, we eliminated the on-site repulsion. Generally the magnitude of $J$ is small in real materials, but we enhance it and study rather artificial model focusing on its influence.

The many-body instabilities of the system are investigated using an unbiased and highly scalable numerical tool, the TUFRG\cite{ref24}. This approach is a recent modification of the functional renormalization group (FRG) method\cite{ref25, ref26, ref27} and closely related to the singular-mode FRG\cite{ref28}, and we now briefly outline the main idea of the approach.

The FRG method is based on the formalism of the quantum field theory where the partition function of the interacting system is expressed as
\begin{equation}\label{eq04}
Z = \int {D\psi D\bar \psi } e^{ - S[\psi ,\bar \psi ]}  = \int {D\psi D\bar \psi } e^{(\bar \psi ,G_0^{ - 1} \psi ) - S_{{\rm int} } [\psi ,\bar \psi ]}.	
\end{equation}
with $\psi ,\bar \psi$ and $S$, being the fermionic Grassmann fields and the action of the system. As a kind of correlation function for the Grassmann fields, the Green function is generated by functional differentiation of the generating functional $W[\eta ,\bar \eta ]$ that is obtained by adding external sources, coupled with the Grassmann fields, into the action $S$. The generating functional of one-particle-irreducible (1PI) vertices, $\Gamma [\psi ,\bar \psi ]$, is then obtained by the Legendre transform of $W[\eta ,\bar \eta]$:
\begin{equation}\label{eq05}
\begin{split}	
W[\eta ,\bar \eta ] &= -\ln \int {D\psi D\bar \psi } e^{ - S[\psi ,\bar \psi ] + (\bar \eta ,\psi ) + (\bar \psi ,\eta )},\\
\Gamma [\psi ,\bar \psi ] &= W[\eta ,\bar \eta ] + (\bar \eta ,\psi ) + (\bar \psi ,\eta ).	
\end{split}	
\end{equation}

To set up the FRG flow equation, the bare propagator $G_0$ in Eq. (\ref{eq04}) is regularized by an infrared cutoff with energy scale $\Omega$, i.e.
\begin{equation}\label{eq06}
G_0 (\omega ,{\bf{k}}) \to G_0^\Omega  (\omega ,{\bf{k}}) = \frac{{\hbar ^2 \omega ^2 }}{{\hbar ^2 \omega ^2  + \Omega ^2 }}G_0 (\omega ,{\bf{k}}).	
\end{equation}
The regularized propagator $G_0^\Omega$ is then used to define the scale-dependent effective action $\Gamma ^\Omega$, which generates the 1PI vertex functions $\gamma ^{(2n),\Omega }$. Taking the derivative of $\Gamma ^\Omega$ with respect to $\Omega$ yields the FRG flow equation, from which an infinite hierarchy of flow equations of the 1PI vertices is derived by Taylor expansion. The initial condition of the FRG flow at ultraviolet scale $\Omega_0$ is given by $\Gamma ^{\Omega _0 }  \equiv \Gamma ^{(0)}  = S_{\rm int}$.

For practical integration of the flow equation, we need to introduce some approximations. First we truncate the hierarchy of the flow equations by discarding all $2n$-point vertices with $n \ge 3$, i.e., by setting as $\gamma ^{(2n),\Omega }  = 0 \quad (n \ge 3)$. Second, the self-energy feedbacks into dressed propagator and 4-point vertices are neglected. Thus, the original infinite hierarchy is reduced to the flow equation only for 4-point vertex $\gamma^{4,\Omega}$. Finally, we neglect the frequency dependence of the vertex $\gamma ^{4,\Omega}$, with only the frequency conservation retained. These approximations have proven to provide reliable results for many two-dimensional (2D) systems\cite{ref25, ref26}.

For spin-SU(2)-invariant systems, the 4-point vertex $\gamma ^{4,\Omega }$ is replaced with the effective interaction $V^\Omega$, and the effective action is represented by it as

\begin{equation}\label{eq07}
\begin{split}	
\Gamma ^\Omega  [\psi ,\bar \psi ] = & \frac{1}{{2N\beta \hbar ^2 }}\sum\limits_{\xi _1 }  \cdots  \sum\limits_{\xi _4 } {V_{o_1 o_2 ,o_3 o_4 }^\Omega  ({\bf{k}}_1 ,{\bf{k}}_2 ;{\bf{k}}_3 ,{\bf{k}}_4 )} \delta _{{\bf{k}}_1  + {\bf{k}}_2 ,{\bf{k}}_3  + {\bf{k}}_4 } \\
& \times \delta _{\omega _1  + \omega _2 ,\omega _3  + \omega _4 } \sum\limits_{\sigma ,\sigma '} {\bar \psi _\sigma  (\xi _1 )\bar \psi _{\sigma '} (\xi _2 )\psi _{\sigma '} (\xi _4 )\psi _\sigma  (\xi _3 )}.
\end{split}
\end{equation}
Here $\xi _i  = (\omega _i ,{\bf{k}}_i ,o_i )$ is multi-index quantum number containing a Matsubara frequency $\omega _i$, wave vector ${\bf{k}}_i$ and sublattice index $o_i$, while $N$ and $\beta$ are the total number of unit cells and the inverse temperature, respectively. With these preliminaries, the evolution of $V^\Omega$ can be derived from the flow equation of $\gamma ^{4,\Omega}$, and it is composed of three contributions.
\begin{equation}\label{eq08}
\frac{d}{{d\Omega }}V^\Omega = J^{{\rm{pp}}} (\Omega ) + J^{{\rm{ph,cr}}} (\Omega ) + J^{{\rm{ph,d}}} (\Omega )
\end{equation}
The concrete expressions of the contributions from the particle-particle ($J^{{\rm{pp}}}$), the crossed particle-hole ($J^{{\rm{ph,cr}}}$) and the direct particle-hole ($J^{{\rm{ph,d}}}$) channels can be found in Ref. \cite{ref29}. By integrating Eq. (\ref{eq08}) one can find the effective interaction:
\begin{equation}\label{eq09}
V^\Omega   = V^{(0)}  + \Phi ^{{\rm{pp}}} (\Omega ) + \Phi ^{{\rm{ph,cr}}} (\Omega ) + \Phi ^{{\rm{ph,d}}} (\Omega ),
\end{equation}
where $V^{(0)}$ is the initial interaction, while the single-channel coupling functions $\Phi ^{{\rm{pp}}} ,\Phi ^{{\rm{ph,cr}}}$ and $\Phi ^{{\rm{ph,d}}}$ are defined by
\begin{equation}\label{eq10}
\begin{split}	
&\Phi ^{{\rm{pp}}} (\Omega ) \equiv \int_{\Omega _0 }^\Omega  {d\Omega '} J^{{\rm{pp}}} (\Omega '),\\
&\Phi ^{{\rm{ph,cr}}} (\Omega ) \equiv \int_{\Omega _0 }^\Omega  {d\Omega '} J^{{\rm{ph,cr}}} (\Omega '),
\Phi ^{{\rm{ph,d}}} (\Omega ) \equiv \int_{\Omega _0 }^\Omega  {d\Omega '} J^{{\rm{ph,d}}} (\Omega ').
\end{split}	
\end{equation}

To achieve a high resolution of the effective interaction in momentum space, Husemann and Salmhofer proposed an efficient parametrization of the single-channel coupling functions\cite{ref30}. In this approach three bosonic propagators are introduced by projections of the three coupling functions onto their own channels:
\begin{equation}\label{eq11}
P^\Omega   = {\rm{\hat P}}[\Phi ^{{\rm{pp}}} (\Omega )],C^\Omega   = {\rm{\hat C}}[\Phi ^{{\rm{ph,cr}}} (\Omega )],
D^\Omega   = {\rm{\hat D}}[\Phi ^{{\rm{ph,d}}} (\Omega )].
\end{equation}
They are matrices that depend only on one transfer momentum, not on three momenta, which reduces greatly the memory required by $V^\Omega$ and makes the approach highly scalable. The projections in Eq. (\ref{eq11}) are performed by means of the plane-wave bases in our work. In numerical implementation, the inverse projections of Eq. (\ref{eq11}) necessitate involving only limited numbers of the bases, and the results can only give approximate values of the coupling functions.
\begin{equation}\label{eq12}
\Phi ^{{\rm{pp}}} (\Omega ) \approx {\rm{\hat P}}^{ - 1} [P^\Omega  ],\Phi ^{{\rm{ph,cr}}} (\Omega ) \approx {\rm{\hat C}}^{ - 1} [C^\Omega  ],\Phi ^{{\rm{ph,d}}} (\Omega ) \approx {\rm{\hat D}}^{ - 1} [D^\Omega  ].
\end{equation}
The detailed representation of the projection and inverse projection is given in Ref. \cite{ref29}. Differentiating Eqs. (\ref{eq10}) and (\ref{eq11}) with respect to $\Omega$, and inserting Eq. (\ref{eq09}) into the expressions for $J^{{\rm{pp}}}, J^{{\rm{ph,cr}}}$ and $J^{{\rm{ph,d}}}$, one can find the flow equation of the bosonic propagators. It contains intricate terms in which internal bosonic propagators appear in the fermionic loops and have to be integrated out, posing a challenge in calculations.

The TUFRG scheme\cite{ref24} introduces additional approximation. With double insertion of truncated projection of unity, it decouples the bosonic propagators from the fermionic one, and the flow equation becomes simplified. Ultimately, the TUFRG flow equation for bosonic propagators is represented as
\begin{equation}\label{eq13}
\begin{split}	
\frac{{dP^\Omega  ({\bf{q}})}}{{d\Omega }} = &V^{{\rm{P}}(\Omega )} ({\bf{q}})\frac{{d\chi ^{{\rm{pp}}} ({\bf{q}})}}{{d\Omega }}V^{{\rm{P}}(\Omega )} ({\bf{q}}),\\
\frac{{dC^\Omega  ({\bf{q}})}}{{d\Omega }} = &V^{{\rm{C}}(\Omega )} ({\bf{q}})\frac{{d\chi ^{{\rm{ph}}} ({\bf{q}})}}{{d\Omega }}V^{{\rm{C}}(\Omega )} ({\bf{q}}),\\
\frac{{dD^\Omega  ({\bf{q}})}}{{d\Omega }} = &[V^{{\rm{C}}(\Omega )} ({\bf{q}}) - V^{{\rm{D}}(\Omega )} ({\bf{q}})]\frac{{d\chi ^{{\rm{ph}}} ({\bf{q}})}}{{d\Omega }}V^{{\rm{D}}(\Omega )} ({\bf{q}})\\
& + V^{{\rm{D}}(\Omega )} ({\bf{q}})\frac{{d\chi ^{{\rm{ph}}} ({\bf{q}})}}{{d\Omega }}[V^{{\rm{C}}(\Omega )} ({\bf{q}}) - V^{{\rm{D}}(\Omega )} ({\bf{q}})].
\end{split}	
\end{equation}
Here $\chi^{{\rm{pp}}}$ and $\chi^{{\rm{ph}}}$ are the susceptibility matrices, and $V^{\rm{P}} (\Omega ),V^{\rm{C}} (\Omega ),V^{\rm{D}} (\Omega)$ are the projections of $V^\Omega$ into three channels, as defined in Eq. (\ref{eq11}). By a combination of Eqs. (\ref{eq09}) and (\ref{eq12}), all the quantities in the flow equation (\ref{eq13}) can be represented via the bosonic propagators, and thus we can obtain a closed system of differential equations expressed only by the propagators. In the case of the plane-wave basis $f_m ({\bf{k}}) = e^{i{\bf{R}}_m \cdot {\bf{k}}}$, the explicit expressions for $V^{\rm{P}} (\Omega ),V^{\rm{C}} (\Omega)$ and $V^{\rm{D}} (\Omega)$, as well as for $\chi ^{{\rm{pp}}}$ and $\chi ^{{\rm{ph}}}$, are presented in Ref. \cite{ref29}.

In numerical implementation, only a limited number of the bases are involved in the computation of the bosonic propagators. Concretely, one can generally set a cut-off radius $R_C$ to be a few times larger than the lattice constant, and then neglect all the elements of the propagators associated with the basis indices $m$ satisfying the condition $|{\bf{R}}_m | > R_C$. We will call these elements \emph{the high lattice harmonic components} of the bosonic propagators. In the case of short-ranged interaction Hamiltonian, this truncation is exact for $V^{{\rm{X}},(0)}  \equiv {\rm{\hat X}}[V^{(0)}]$ (${\rm{X}}={\rm{P}}$, C, or D). But the projections of the effective interaction may have non-negligible high lattice (rapidly varying in momentum space) harmonics. From eqs. (\ref{eq09}) and (\ref{eq12}), one can derive, e.g., $V^{\rm{P}} (\Omega ) = V^{{\rm{P}},(0)}  + P^\Omega   + V^{{\rm{P}} \leftarrow {\rm{C}}} (\Omega ) + V^{{\rm{P}} \leftarrow {\rm{D}}} (\Omega )$. Although the aforementioned truncation is valid for $V^{{\rm{P}},(0)}$ and $P^\Omega$, the two crossed contributions, $V^{{\rm{P}} \leftarrow {\rm{C}}} (\Omega ) \equiv {\rm{\hat P}}\{ {\rm{\hat C}}^{ - 1} [C^\Omega]\}$ and $V^{{\rm{P}} \leftarrow {\rm{D}}} (\Omega ) \equiv {\rm{\hat P}}\{ {\rm{\hat D}}^{ - 1} [D^\Omega  ]\}$, could have considerable high harmonic components, leading to slow convergence in the expansion $V^\Omega   = {\rm{\hat P}}^{ - 1} [V^{\rm{P}} (\Omega )]$. In particular, this tendency gets more apparent when multiple orders interact with each other. Therefore the projection errors from neglecting the high lattice harmonic components may affect the reliability of the results, especially in the case of coexistence phases or in the vicinity of phase boundaries. However, we note that a previous study\cite{ref40} has shown that only the low lattice (slowly varying in momentum space) harmonic components contribute significantly to the electronic instabilities. In general, it is expected that the advantage from a higher momentum resolution of $V^\Omega$ in the TUFRG would win over its drawback from the truncation in the expansion for $V^\Omega$.

Due to its simplified structure, the TUFRG is known to ensure a fast and highly resolved computation and has been successfully applied to the analysis of the electronic instabilities in various 2D one-band\cite{ref24, ref31} and multi-band systems\cite{ref15, ref20, ref29, ref32}, and even in three-dimensional system\cite{ref33}. Recently, it has been extended to address more complicated systems\cite{ref34, ref35, ref36}.

\begin{figure}[h!]
	\begin{center}
		\includegraphics[width=8.0cm]{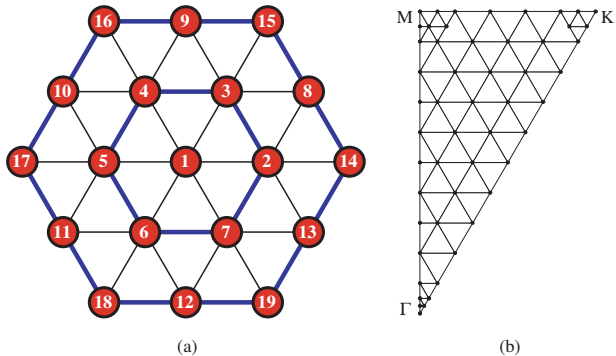}
	\end{center}
	\caption{(Color online) (a) Bravais lattice vectors ${\bf{R}}_1 , \cdots ,{\bf{R}}_{N_b } (N_b  = 19)$ of the plain-wave bases $f_m ({\bf{p}}) = e^{i{\bf{R}}_m  \cdot {\bf{p}}}$ used in our TUFRG calculation. All the bosonic propagators containing the basis indices larger than $N_b$ are neglected. (b) Mesh of the sampling points for transfer momenta within the irreducible region of the BZ. The points are distributed more densely near the $\Gamma,\rm {K}$ and $\rm {M}$ points. The bosonic propagators are numerically calculated only for these sampling points.}
	\label{fig1}
\end{figure}

For our implementation we used 19 plain-wave form-factor bases ($N_b  = 19$) with Bravais lattice vectors lying within second hexagonal shell on the triangular lattice as shown in Fig. \ref{fig1}(a). The choice of $N_b  = 19$ has been justified by a convergence test with respect to the form-factor bases in Ref. \cite{ref31}. The TUFRG flow equation (\ref{eq13}) is solved only for sampling transfer momenta in the irreducible region of the Brillouin zone (BZ). The mesh of those transfer momenta ($\bf q$ mesh) is shown in Fig. \ref{fig1}(b).

In each step of integration of the flow equation, the point-group symmetry relations\cite{ref20} and the filtering process\cite{ref29} are employed to generate the bosonic propagators outside of the region, thus reducing the numerical effort by a factor of twelve. The initial values of the projections, $V^{\rm{P}} (\Omega ),V^{\rm{C}} (\Omega )$ and $V^{\rm{D}} (\Omega )$, are determined by projecting the interaction Hamiltonian in momentum space onto the three channels\cite{ref20}. The diverse ordering tendencies are analyzed on a qualitative level by means of the linear-response-based approach for identifying the type of order\cite{ref20}.

\begin{figure}[h!]
	\begin{center}
		\includegraphics[width=8.0cm]{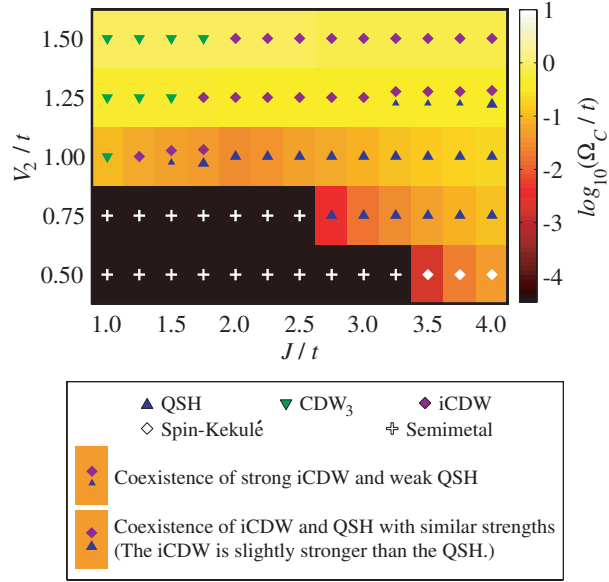}
	\end{center}
	\caption{(Color online) Schematic ground-state phase diagram in the space of parameters $V_2$ and $J$. The color bar indicates values of critical scales $\Omega_C$ that can be regarded as the estimates for the transition temperatures. All the notations have the same meaning as those in Ref. \cite{ref20}. The coexistence phases are identified via the same criterion as in Ref. \cite{ref29}.}
	\label{fig2}
\end{figure}

We have scanned the region of the parameter space, $V_2  = 0.5t\sim 1.5t$ and $J = t\sim 4t$, which covers the entire part of the QSH phase in the phase diagram of Ref. \cite{ref20}. The results are summarized in the schematic phase diagram shown in Fig. \ref{fig2}. The critical scales $\Omega_C$, at which some bosonic propagators get divergent, are also provided using the color bar. The scale $\Omega_C$ can be interpreted as an estimate for the transition temperature. The phase diagram includes the QSH (QSH), the three-sublattice charge-density-wave ($\rm {CDW}_3$), the incommensurate charge-density-wave (iCDW), and the spin-Kekul\'{e} (spin-Kekul\'{e}) phases.

Main structure of the resulting phase diagram is quite similar to the one in Ref. \cite{ref20}. The QSH phase is observed in a small interval $0.75t \le V_2  \le 1.25t$. A spin bond-ordered state, dubbed spin-Kekul\'{e}, emerges in the region of large $J$ and small $V_2$. For small $J$ and large $V_2$, the $\rm {CDW}_3$ phase appears. When increasing $J$, it turns into the iCDW phase. A considerable increase in critical scale is observed upon increasing $V_2$. The ordering tendencies are sensitive to the parameter $V_2$, but not to $J$. The schematic patterns of the corresponding order parameters for the QSH, spin-Kekul\'{e} and $\rm{CDW}_3$ can be found in Ref. \cite{ref20}.

In the region of the semimetal, there is no divergence of any bosonic propagator observed in the TUFRG flow down to the stopping scale $\Omega ^*  = 4.8 \times 10^{ - 5} {\rm{eV}}$. The coexistence phases are identified via the same criterion as in Ref. \cite{ref29}. Concretely, the notation \emph{Coexistence of strong iCDW and weak QSH} means that, at the critical scale, the most positive eigenvalue $\lambda _{{\rm{iCDW}}}$, of the effective charge-susceptibility matrix $W^{{\rm{CHG}}} ({\bf{Q}}_{{\rm{iCDW}}})$, is 2--5 times larger than that of the effective spin-susceptibility matrix $W^{{\rm{SPN}}} ({\bf{Q}} = 0)$, $\lambda _{{\rm{QSH}}}$, while the notation \emph{Coexistence of iCDW and QSH with similar strengths} means the relation $\lambda _{{\rm{iCDW}}} /2 < \lambda _{{\rm{QSH}}}  \le \lambda _{{\rm{iCDW}}}$.

\section{Combination of TUFRG and MF theory}\label{sec3}

From now on, we will restrict our consideration to the QSH phase with the parameter setting $V_2=t$ and $J=2t$. In this parameter setting, the calculated bosonic propagators at the critical scale $\Omega_C$ exhibit divergences only in the particle-hole channels, as represented by
\begin{equation}\label{eq14}
\begin{split}	
&P^{\Omega _C } ({\bf{q}}) \approx 0,C^{\Omega _C } ({\bf{q}}) \approx C_{{\rm{singul}}}^{\Omega _C } ({\bf{q}} = 0)\delta _{{\bf{q}},0},\\
&D^{\Omega _C } ({\bf{q}}) \approx D_{{\rm{singul}}}^{\Omega _C } ({\bf{q}} = 0)\delta _{{\bf{q}},0}  = \frac{1}{2}C_{{\rm{singul}}}^{\Omega _C } ({\bf{q}} = 0)\delta _{{\bf{q}},0},
\end{split}	
\end{equation}
with the singular part $C_{{\rm{singul}}}^{\Omega _C }$ of the matrix $C^{\Omega _C}$ having an unique singular eigenmode $\phi$,
\begin{equation}\label{eq15}
[C_{{\rm{singul}}}^{\Omega _C } ({\bf{q}} = 0)]_{o_1 o_2 m,o_3 o_4 n}  = \lambda _C (\phi _{o_1 o_2 m} )(\phi _{o_3 o_4 n} )^* .
\end{equation}
The value of $\lambda_C$ is $\lambda _C  = 729.563t$, and the values of the elements of the singular mode $\phi$ are given in Table \ref{tab01}.

\begin{table}[h!]
	\caption{The values of the elements $\phi _{oo'm}$ of the unique singular mode for the bosonic propagator $C^{\Omega _C } ({\bf{q}} = 0)$. Unspecified elements are zero. The values of constants here are $g_1=0.28827$ and $g_2=0.01529$, respectively.}
	\begin{center}
		\begin{tabular}{|c|c|c|c|c|c|c|}
			\hline
			$oo'm$ & $AA2$ & $AA3$ & $AA4$ & $AA5$ & $AA6$ & $AA7$\\
			\hline
			$\phi_{oo'm}$ & $ig_1$ & $-ig_1$ & $ig_1$ & $-ig_1$ & $ig_1$ & $-ig_1$\\
			\hline
			$oo'm$ & $AA14$ & $AA15$ & $AA16$ & $AA17$ & $AA18$ & $AA19$\\
			\hline
			$\phi_{oo'm}$ & $-ig_2$ & $ig_2$ & $-ig_2$ & $ig_2$ & $-ig_2$ & $ig_2$\\
			\hline
			$oo'm$ & $BB2$ & $BB3$ & $BB4$ & $BB5$ & $BB6$ & $BB7$\\
			\hline
			$\phi_{oo'm}$ & $-ig_1$ & $ig_1$ & $-ig_1$ & $ig_1$ & $-ig_1$ & $ig_1$\\
			\hline
			$oo'm$ & $BB14$ & $BB15$ & $BB16$ & $BB17$ & $BB18$ & $BB19$\\
			\hline
			$\phi_{oo'm}$ & $ig_2$ & $-ig_2$ & $ig_2$ & $-ig_2$ & $ig_2$ & $-ig_2$\\
			\hline
		\end{tabular}
	\end{center}
	\label{tab01}
\end{table}

Eqs. (\ref{eq14}) and (\ref{eq15}) dictate that the effective action takes the following form (for details, see Ref. \cite{ref29}):
\begin{equation}\label{eq16}
\begin{split}	
\Gamma ^{\Omega _C } [\psi ,\bar \psi ] \approx &\frac{1}{{2N\beta \hbar ^2 }}\sum\limits_{{\bf{k}},{\bf{p}}} {\sum\limits_{\omega _1  \cdots \omega _4 } {\sum\limits_{o_1  \cdots o_4 } {\sum\limits_{m,n} {\lambda _C (\phi _{o_1 o_2 m} )(\phi _{o_3 o_4 n} )^* f_m^* ({\bf{p}})f_n ({\bf{k}})} } } } \\
&\times \delta _{\omega _1  + \omega _4 ,\omega _2  + \omega _3 } \sum\limits_{\sigma ,\sigma '} {\bar \psi _\sigma  ({\bf{p}},\omega _1 ,o_1 )\bar \psi _{\sigma '} ({\bf{k}},\omega _4 ,o_4 )\psi _{\sigma '} ({\bf{p}},\omega _2 ,o_2 )\psi _\sigma  ({\bf{k}},\omega _3 ,o_3 )} \\ 
+ &\frac{1}{{2N\beta \hbar ^2 }}\sum\limits_{{\bf{k}},{\bf{p}}} {\sum\limits_{\omega _1  \cdots \omega _4 } {\sum\limits_{o_1  \cdots o_4 } {\sum\limits_{m,n} {\frac{1}{2}\lambda _C (\phi _{o_1 o_2 m} )(\phi _{o_3 o_4 n} )^* f_m^* ({\bf{p}})f_n ({\bf{k}})} } } } \\
&\times \delta _{\omega _1  + \omega _4 ,\omega _2  + \omega _3 } \sum\limits_{\sigma ,\sigma '} {\bar \psi _\sigma  ({\bf{p}},\omega _1 ,o_1 )\bar \psi _{\sigma '} ({\bf{k}},\omega _4 ,o_4 )\psi _{\sigma '} ({\bf{k}},\omega _3 ,o_3 )\psi _\sigma  ({\bf{p}},\omega _2 ,o_2 )}.
\end{split}	
\end{equation}
From the relation,
\begin{equation}\label{eq17}
\begin{split}	
\sum\limits_{\sigma ,\sigma '} &{\bar \psi _\sigma ({\bf{p}},\omega _1 ,o_1 )} \bar \psi _{\sigma '} ({\bf{k}},\omega _4 ,o_4 )\psi _{\sigma '} ({\bf{p}},\omega _2 ,o_2 )\psi _\sigma  ({\bf{k}},\omega _3 ,o_3 ) \\
= &-\frac{1}{2}\left[ {\sum\limits_s {\bar \psi _s ({\bf{p}},\omega _1 ,o_1 )\psi _s ({\bf{p}},\omega _2 ,o_2 )} } \right]\left[ {\sum\limits_\sigma  {\bar \psi _\sigma  ({\bf{k}},\omega _4 ,o_4 )\psi _\sigma  ({\bf{k}},\omega _3 ,o_3 )} } \right] \\
& - \frac{1}{2}\left[ {\sum\limits_{s,s'} {\bar \psi _s ({\bf{p}},\omega _1 ,o_1 )\vec \sigma _{ss'} \psi _{s'} ({\bf{p}},\omega _2 ,o_2 )} } \right] \cdot \left[ {\sum\limits_{\sigma ,\sigma '} {\bar \psi _\sigma  ({\bf{k}},\omega _4 ,o_4 )\vec \sigma _{\sigma \sigma '} \psi _{\sigma '} ({\bf{k}},\omega _3 ,o_3 )} } \right],
\end{split}	
\end{equation}
we have
\begin{equation}\label{eq18}
\begin{split}	
\Gamma ^{\Omega _C } [\psi ,\bar \psi ] \approx & - \frac{1}{2}\frac{1}{{2N\beta \hbar ^2 }}\sum\limits_{{\bf{k}},{\bf{p}}} {\sum\limits_{\omega _1  \cdots \omega _4 } {\sum\limits_{o_1  \cdots o_4 } {\sum\limits_{m,n} {\lambda _C (\phi _{o_1 o_2 m} )(\phi _{o_3 o_4 n} )^* f_m^* ({\bf{p}})f_n ({\bf{k}})} } } } \delta _{\omega _1  + \omega _4 ,\omega _2  + \omega _3 } \\
& \times \left[ {\sum\limits_{s,s'} {\bar \psi _s ({\bf{p}},\omega _1 ,o_1 )\vec \sigma _{ss'} \psi _{s'} ({\bf{p}},\omega _2 ,o_2 )} } \right] \cdot \left[ {\sum\limits_{\sigma ,\sigma '} {\bar \psi _\sigma  ({\bf{k}},\omega _4 ,o_4 )\vec \sigma _{\sigma \sigma '} \psi _{\sigma '} ({\bf{k}},\omega _3 ,o_3 )} } \right],
\end{split}	
\end{equation}
which implies a strong spin-spin (magnetic) interaction and a possible instability in the spin channel.

On the other hand, the TUFRG flow should be stopped at the critical scale because the truncation of the hierarchy of the flow equations is no longer justified. To complete the calculation, we can use the MF theory based on the calculated vertex functions. The effective action in Eq. (\ref{eq18}) is equivalent to the following effective Hamiltonian,
\begin{equation}\label{eq19}
\begin{split}	
H^{\Omega _C }  = &  - \frac{1}{2}\frac{1}{{2N}}\sum\limits_{{\bf{k}},{\bf{p}}} {\sum\limits_{o_1  \cdots o_4 } {\sum\limits_{m,n} {\lambda _C (\phi _{o_1 o_2 m} )(\phi _{o_3 o_4 n} )^* f_m^* ({\bf{p}})f_n ({\bf{k}})} } } \\
& \times \left[ {\sum\limits_{s,s'} {c_{{\bf{p}},o_1 ,s}^ \dagger  \vec \sigma _{ss'} c_{{\bf{p}},o_2 ,s'} } } \right] \cdot \left[ {\sum\limits_{\sigma ,\sigma '} {c_{{\bf{k}},o_4 ,\sigma }^ \dagger  \vec \sigma _{\sigma \sigma '} c_{{\bf{k}},o_3 ,\sigma '} } } \right].
\end{split}	
\end{equation}
One can simply use it as an input for the MF calculation, but it would account doubly the high-energy modes, leading to an overestimation of the ordering tendencies. In this paper we adopt an idea of Wang, Eberlein and Metzner\cite{ref37}, in which only the irreducible part of the 4-point vertex enters the MF equation.

The TUFRG flow equation for bosonic propagator in the spin channel reads
\begin{equation}\label{eq20}
\frac{{dC^\Omega  ({\bf{q}})}}{{d\Omega }} = V^{{\rm{C}}(\Omega )} ({\bf{q}})\frac{{d\chi ^{{\rm{ph}}} ({\bf{q}})}}{{d\Omega }}V^{{\rm{C}}(\Omega )} ({\bf{q}}).
\end{equation}
Following Wang {\it{et al}}.\cite{ref37}, we compute the propagator by integrating the above flow equation at the scale $\Omega  > \Omega _C$. At the critical scale, the projection matrix $V^{{\rm{C}}(\Omega _C )}$ of the effective interaction consists of the dominant part $C^{\Omega _C }$ and two other crossed contributions, $V^{{\rm{C}} \leftarrow {\rm{P}}} (\Omega _C )$ and $V^{{\rm{C}} \leftarrow {\rm{D}}} (\Omega _C )$\cite{ref29}. Below $\Omega _C$, we will neglect the effect of the crossed contributions and employ an approximation $V^{{\rm{C}}(\Omega )}  \approx C^\Omega$. Thus, at lower scale $\Omega  < \Omega _C$, the flow equation (\ref{eq20}) becomes
\begin{equation}\label{eq21}
\frac{{dC^\Omega  ({\bf{q}})}}{{d\Omega }} = C^\Omega  ({\bf{q}})\frac{{d\chi ^{{\rm{ph}}} ({\bf{q}})}}{{d\Omega }}C^\Omega  ({\bf{q}}) \hspace{2pc} \textrm {for } \Omega < \Omega_C ,
\end{equation}
leading to the exact solution,
\begin{equation}\label{eq22}
[C^\Omega  ({\bf{q}})]^{ - 1}  - [C^{\Omega _C } ({\bf{q}})]^{ - 1}  = \chi ^{{\rm{ph}}(\Omega _C )} ({\bf{q}}) - \chi ^{{\rm{ph}}(\Omega )} ({\bf{q}}) \hspace{2pc} \textrm {for } \Omega  < \Omega _C.
\end{equation}
Now we introduce the irreducible bosonic propagator $\tilde C({\bf{q}})$ defined by
\begin{equation}\label{eq23}
[\tilde C({\bf{q}})]^{ - 1}  = [C^{\Omega _C } ({\bf{q}})]^{ - 1}  + \chi ^{{\rm{ph}}(\Omega _C )} ({\bf{q}}) .
\end{equation}
Eq. (\ref{eq22}) can be expressed as
\begin{equation}\label{eq24}
[C^\Omega  ({\bf{q}})]^{ - 1}  = [\tilde C({\bf{q}})]^{ - 1}  - \chi ^{{\rm{ph}}(\Omega )} ({\bf{q}}) \hspace{2pc} \textrm {for } \Omega  < \Omega _C ,
\end{equation}
which is nothing but the random phase approximation (RPA) in the spin channel.

It is well known that the RPA in a given channel has a critical condition identical to that in the MF theory for the same channel. So we suggest taking $\tilde C({\bf{q}})$, not $C^{\Omega _C } ({\bf{q}})$, as an input for the MF calculation in the spin channel. More explicitly, in the present case, we will use
\begin{equation}\label{eq25}
\begin{split}	
H_{{\rm{irred}}}  = & - \frac{1}{2}\frac{1}{{2N}}\sum\limits_{{\bf{k}},{\bf{p}}} {\sum\limits_{o_1  \cdots o_4 } {\sum\limits_{m,n} {[\tilde C({\bf{q}} = 0)]_{o_1 o_2 m,o_3 o_4 n} f_m^* ({\bf{p}})f_n ({\bf{k}})} } } \\
& \times \left[ {\sum\limits_{s,s'} {c_{{\bf{p}},o_1 ,s}^ \dagger  \vec \sigma _{ss'} c_{{\bf{p}},o_2 ,s'} } } \right] \cdot \left[ {\sum\limits_{\sigma ,\sigma '} {c_{{\bf{k}},o_4 ,\sigma }^ \dagger  \vec \sigma _{\sigma \sigma '} c_{{\bf{k}},o_3 ,\sigma '} } } \right]
\end{split}	
\end{equation}
as an input interaction Hamiltonian for the MF theory.

From Eqs. (\ref{eq14}) and (\ref{eq15}) we have $C^{\Omega _C } ({\bf{q}} = 0) \approx C_{{\rm{singul}}}^{\Omega _C } ({\bf{q}} = 0) = \lambda _C \left| \phi  \right\rangle \left\langle \phi  \right| $, and inserting it into Eq. (\ref{eq23}), we get the following result,
\begin{equation} \nonumber
\begin{split}	
\tilde C({\bf{q}} = 0) & = \sqrt {C^{\Omega _C } ({\bf{q}} = 0)} \\
& \times \left( {1 + \sqrt {C^{\Omega _C } ({\bf{q}} = 0)} \chi ^{{\rm{ph}}(\Omega _C )} ({\bf{q}} = 0)\sqrt {C^{\Omega _C } ({\bf{q}} = 0)} } \right)^{ - 1} \sqrt {C^{\Omega _C } ({\bf{q}} = 0)} \\
& = \sqrt {\lambda _C } \left[ {1 + \sqrt {\lambda _C } \left( {\left\langle \phi  \right|\chi ^{{\rm{ph}}(\Omega _C )} ({\bf{q}} = 0)\left| \phi  \right\rangle } \right)\sqrt {\lambda _C } } \right]^{ - 1} \sqrt {\lambda _C } \left| \phi  \right\rangle \left\langle \phi  \right|.
\end{split}	
\end{equation}
Thus the irreducible bosonic propagator is given by
\begin{equation}\label{eq26}
\tilde C({\bf{q}} = 0) = \tilde \lambda _C \left| \phi  \right\rangle \left\langle \phi  \right|,
\end{equation}
where the irreducible coupling constant $\tilde \lambda _C$ is defined by
\begin{equation}\label{eq27}
\tilde \lambda _C  \equiv \lambda _C \left[ {1 + \lambda _C \overline {\chi ^{{\rm{ph}}} } } \right]^{ - 1} ,
\end{equation}
with
\begin{equation}\label{eq28}
\begin{split}	
\overline {\chi ^{{\rm{ph}}} } & = \left\langle \phi  \right|\chi ^{{\rm{ph}}(\Omega _C )} ({\bf{q}} = 0)\left| \phi  \right\rangle \\
& = \sum\limits_{o_1 o_2 m} {\sum\limits_{o_3 o_4 n} {(\phi _{o_1 o_2 m} )^* } } [\chi ^{{\rm{ph}}(\Omega _C )} ({\bf{q}} = 0)]_{o_1 o_2 m,o_3 o_4 n} (\phi _{o_3 o_4 n} ).
\end{split}	
\end{equation}
From the values of $\lambda _C  = 729.563t$ and $\overline {\chi ^{{\rm{ph}}} }  = 0.764t^{ - 1}$ we obtain $\tilde \lambda _C  = 1.307t$.

Now we consider the following interaction Hamiltonian,
\begin{equation}\label{eq29}
\begin{split}	
H_{{\rm{irred}}}  = & - \frac{1}{2}\frac{1}{{2N}}\tilde \lambda _C \left[ {\sum\limits_{o_1 o_2 m} {\sum\limits_{\bf{p}} {(\phi _{o_1 o_2 m} )^* f_m ({\bf{p}})} } \sum\limits_{s,s'} {c_{{\bf{p}},o_2 ,s}^ \dagger  \vec \sigma _{ss'} c_{{\bf{p}},o_1 ,s'} } } \right]^ \dagger \\
& \cdot \left[ {\sum\limits_{o_3 o_4 n} {\sum\limits_{\bf{k}} {(\phi _{o_3 o_4 n} )^* f_n ({\bf{k}})} } \sum\limits_{\sigma ,\sigma '} {c_{{\bf{k}},o_4 ,\sigma }^ \dagger  \vec \sigma _{\sigma \sigma '} c_{{\bf{k}},o_3 ,\sigma '} } } \right].
\end{split}	
\end{equation}
In the MF theory it is approximated as
\begin{equation}\label{eq30}
\begin{split}	
H_{{\rm{irred}}} & \approx H_{{\rm{irred}}}^{{\rm{MF}}}  =  - \frac{1}{2}\vec \Delta  \cdot \left[ {\sum\limits_{oo'm} {\sum\limits_{\bf{k}} {(\phi _{oo'm} )^* f_m ({\bf{k}})} } \sum\limits_{\sigma ,\sigma '} {c_{{\bf{k}},o',\sigma }^ \dagger  \vec \sigma _{\sigma \sigma '} c_{{\bf{k}},o,\sigma '} } } \right]^ \dagger  \\
& - \frac{1}{2}\vec \Delta ^*  \cdot \left[ {\sum\limits_{oo'm} {\sum\limits_{\bf{k}} {(\phi _{oo'm} )^* f_m ({\bf{k}})} } \sum\limits_{\sigma ,\sigma '} {c_{{\bf{k}},o',\sigma }^ \dagger  \vec \sigma _{\sigma \sigma '} c_{{\bf{k}},o,\sigma '} } } \right] + N\frac{{|\vec \Delta |^2 }}{{\tilde \lambda _C }} .
\end{split}	
\end{equation}
Here the vector $\vec \Delta$ is defined by
\begin{equation}\label{eq31}
\vec \Delta  \equiv \frac{{\tilde \lambda _C }}{{2N}}\left\langle {\sum\limits_{oo'm} {\sum\limits_{\bf{k}} {(\phi _{oo'm} )^* f_m ({\bf{k}})} } \sum\limits_{\sigma ,\sigma '} {c_{{\bf{k}},o',\sigma }^ \dagger  \vec \sigma _{\sigma \sigma '} c_{{\bf{k}},o,\sigma '} } } \right\rangle .
\end{equation}
Taking into account the relation $\phi _{oo'm}  = (\phi _{o',o, - {\bf{R}}_m } )^* $ satisfied by the eigenmode $\phi$ in Table \ref{tab01}, one can easily verify that the operator
\begin{equation} \nonumber
A \equiv \sum\limits_{oo'm} {\sum\limits_{\bf{k}} {(\phi _{oo'm} )^* f_m ({\bf{k}})} } \sum\limits_{\sigma ,\sigma '} {c_{{\bf{k}},o',\sigma }^ \dagger  \vec \sigma _{\sigma \sigma '} c_{{\bf{k}},o,\sigma '} }
\end{equation}
is Hermitian, i.e., $A = A^ \dagger$. As a consequence, the vector $\vec \Delta$ should be real valued and Eq. (\ref{eq30}) becomes
\begin{equation} \nonumber
H_{{\rm{irred}}}^{{\rm{MF}}}  =  - \vec \Delta  \cdot \left[ {\sum\limits_{oo'm} {\sum\limits_{\bf{k}} {(\phi _{oo'm} )^* f_m ({\bf{k}})} } \sum\limits_{\sigma ,\sigma '} {c_{{\bf{k}},o',\sigma }^ \dagger  \vec \sigma _{\sigma \sigma '} c_{{\bf{k}},o,\sigma '} } } \right] + N\frac{{|\vec \Delta |^2 }}{{\tilde \lambda _C }} .
\end{equation}
Furthermore, due to the spin-rotation invariance of the system, we can take, without loss of generality, the vector $\vec \Delta$ to be directed along the $z$-axis, so that $S^z$ is conserved, and spin-up and spin-down electrons decouple. Then we have the following interaction Hamiltonian and self-consistency condition,
\begin{equation}\label{eq32}
H_{{\rm{irred}}}^{{\rm{MF}}}  =  - \Delta \left[ {\sum\limits_{oo'm} {\sum\limits_{\bf{k}} {(\phi _{oo'm} )^* f_m ({\bf{k}})} } \sum\limits_{\sigma  = 1, - 1} {\sigma c_{{\bf{k}},o',\sigma }^ \dagger  c_{{\bf{k}},o,\sigma } } } \right] + N\frac{{\Delta ^2 }}{{\tilde \lambda _C }},
\end{equation}
\begin{equation}\label{eq33}
\Delta  = \frac{{\tilde \lambda _C }}{{2N}}\left\langle {\sum\limits_{oo'm} {\sum\limits_{\bf{k}} {(\phi _{oo'm} )^* f_m ({\bf{k}})} } \sum\limits_{\sigma  = 1, - 1} {\sigma c_{{\bf{k}},o',\sigma }^ \dagger  c_{{\bf{k}},o,\sigma } } } \right\rangle .
\end{equation}
Inserting values of $\phi _{oo'm}$ presented in Table \ref{tab01} into Eqs. (\ref{eq32}) and (\ref{eq33}), we get more detailed expressions for $H_{{\rm{irred}}}^{{\rm{MF}}}$ and $\Delta$,
\begin{equation}\label{eq34}
\begin{split}	
&H_{{\rm{irred}}}^{{\rm{MF}}}  = h_{\sigma  = 1}  + h_{\sigma  =  - 1}  + N\frac{{\Delta ^2 }}{{\tilde \lambda _C }}, \\
&h_\sigma   \equiv  - \Delta \sigma \sum\limits_{\bf{k}} {\eta ({\bf{k}})} (c_{{\bf{k}},A,\sigma }^ \dagger  c_{{\bf{k}},A,\sigma }  - c_{{\bf{k}},B,\sigma }^ \dagger  c_{{\bf{k}},B,\sigma } ),
\end{split}	
\end{equation}
\begin{equation}\label{eq35}
\Delta  = \frac{{\tilde \lambda _C }}{{2N}}\sum\limits_{\bf{k}} {\eta ({\bf{k}})\sum\limits_{\sigma  = \pm 1} \sigma  \left( {\left\langle {c_{{\bf{k}},A,\sigma }^ \dagger  c_{{\bf{k}},A,\sigma } } \right\rangle  - \left\langle {c_{{\bf{k}},B,\sigma }^ \dagger  c_{{\bf{k}},B,\sigma } } \right\rangle } \right)} .
\end{equation}
Here the function $\eta ({\bf{k}})$ is defined as
\begin{equation}\label{eq36}
\begin{split}	
\eta ({\bf{k}}) \equiv & 2g_1 [\sin (k_x a) + \sin ( - k_x a/2 + \sqrt 3 k_y a/2) \\
& \hspace{3pc} + \sin ( - k_x a/2 - \sqrt 3 k_y a/2)] \\
& - 2g_2 [\sin (2k_x a) + \sin ( - k_x a + \sqrt 3 k_y a) + \sin ( - k_x a - \sqrt 3 k_y a)] ,
\end{split}	
\end{equation}
with the lattice constant (the distance between next-nearest-neighbor sites) $a$ and the constants, $g_1  = {\rm{0}}{\rm{.28827}}$ and $g_2  = {\rm{0}}{\rm{.01529}}$.

On the other hand, the single-particle Hamiltonian in Eq. (\ref{eq02}) is represented in the momentum space as
\begin{equation}\label{eq37}
\begin{split}	
&H_0  =  - t\sum\limits_{{\bf{k}},\sigma } {\sum\limits_{o,o'} {c_{{\bf{k}},o,\sigma }^ \dagger  } } \left[ {({\mathop{\rm Re}\nolimits} F({\bf{k}}))\sigma ^x  - ({\mathop{\rm Im}\nolimits} F({\bf{k}}))\sigma ^y } \right]_{oo'} c_{{\bf{k}},o',\sigma } ,\\
&F({\bf{k}}) \equiv 1 + 2\cos \left( {\frac{1}{2}k_x a} \right)e^{ - i\frac{{\sqrt 3 }}{2}k_y a} .
\end{split}	
\end{equation}
Finally, we get the following total MF Hamiltonian:
\begin{equation}\label{eq38}
\begin{split}	
H^{{\rm{MF}}}  =& H_0  + H_{{\rm{irred}}}^{{\rm{MF}}}  = H_{\sigma  = 1}^{{\rm{MF}}}  + H_{\sigma  =  - 1}^{{\rm{MF}}}  + N\frac{{\Delta ^2 }}{{\tilde \lambda _C }} , \\
H_\sigma ^{{\rm{MF}}}  \equiv&  - \sum\limits_{\bf{k}} {\sum\limits_{o,o'} {c_{{\bf{k}},o,\sigma }^ \dagger  } } \left[ {t({\mathop{\rm Re}\nolimits} F({\bf{k}}))\sigma ^x  - t({\mathop{\rm Im}\nolimits} F({\bf{k}}))\sigma ^y  + \Delta \sigma \eta ({\bf{k}})\sigma ^z } \right]_{oo'} c_{{\bf{k}},o',\sigma } .
\end{split}	
\end{equation}

\section{Analysis of QSH state}\label{sec4}

\begin{figure}[h!]
	\begin{center}
		\includegraphics[width=8.0cm]{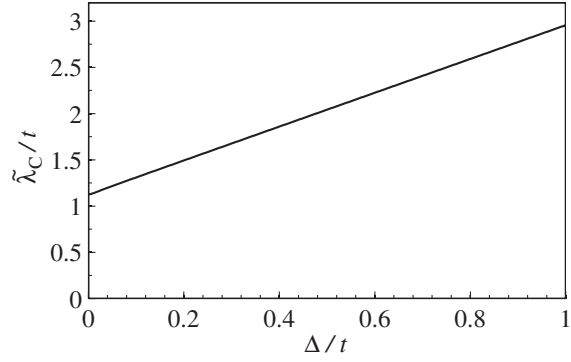}
	\end{center}
	\caption{Relation between $\tilde \lambda _C /t$ and $\Delta /t$ determined by the self-consistency condition. The coupling parameter $\Delta$ is nearly linearly increased with the irreducible coupling constant $\tilde \lambda _C$.}
	\label{fig3}
\end{figure}

As can be seen from the spin-decoupled Hamiltonian in Eq. (\ref{eq38}), the spin-up and spin-down electrons constitute the two-band systems separately. The Hamiltonian can be expressed as
\begin{equation}\label{eq39}
\begin{split}	
H_\sigma ^{{\rm{MF}}}  = \sum\limits_{\bf{k}} {\sum\limits_{o,o'} {c_{{\bf{k}},o,\sigma }^ \dagger  } } \left[ {{\bf{d}}_\sigma  ({\bf{k}}) \cdot {\vec{\sigma}} } \right]_{oo'} c_{{\bf{k}},o',\sigma }, \\
{\bf{d}}_{\sigma  =  \pm 1} ({\bf{k}}) \equiv \left( { - t{\mathop{\rm Re}\nolimits} F({\bf{k}}),t{\mathop{\rm Im}\nolimits} F({\bf{k}}), \mp \Delta \eta ({\bf{k}})} \right).
\end{split}	
\end{equation}
It is straightforward to diagonalize the two-band model to find the dispersion relation and the expectation values of some quantities. Namely, the model has eigenvalues, $\pm |{\bf{d}}_\sigma  ({\bf{k}})|$. At the half-filling, the upper band with the states $\left| {\psi _\sigma ^ +  ({\bf{k}})} \right\rangle$ at the energies $|{\bf{d}}_\sigma  ({\bf{k}})|$ is empty, while the lower band with the states $\left| {\psi _\sigma ^ -  ({\bf{k}})} \right\rangle$ at $-|{\bf{d}}_\sigma  ({\bf{k}})|$ is fully filled. One can easily derive the following relation,
\begin{equation}\label{eq40}
\begin{split}	
\left\langle {c_{{\bf{k}},A,\sigma }^ \dagger  c_{{\bf{k}},A,\sigma } } \right\rangle  & = \left\langle {\psi _\sigma ^ -  ({\bf{k}})} \right|c_{{\bf{k}},A,\sigma }^ \dagger  c_{{\bf{k}},A,\sigma } \left| {\psi _\sigma ^ -  ({\bf{k}})} \right\rangle  \\
&= \frac{1}{2}\left( {1 + \frac{{\sigma \Delta \eta ({\bf{k}})}}{{\sqrt {t^2 |F({\bf{k}})|^2  + \Delta ^2 \eta ^2 ({\bf{k}})} }}} \right), \\
\left\langle {c_{{\bf{k}},B,\sigma }^ \dagger  c_{{\bf{k}},B,\sigma } } \right\rangle  &= \frac{1}{2}\left( {1 - \frac{{\sigma \Delta \eta ({\bf{k}})}}{{\sqrt {t^2 |F({\bf{k}})|^2  + \Delta ^2 \eta ^2 ({\bf{k}})} }}} \right).
\end{split}	
\end{equation}
Inserting this relation into Eq. (\ref{eq35}) yields the following self-consistency condition:
\begin{equation}\label{eq41}
\frac{{\tilde \lambda _C }}{t} = \left[ {\int_{{\rm{BZ}}} {\frac{{d^2 k}}{{S_{{\rm{BZ}}} }}} \frac{{\eta ^2 ({\bf{k}})}}{{\sqrt {|F({\bf{k}})|^2  + (\Delta /t)^2 \eta ^2 ({\bf{k}})} }}} \right]^{ - 1},
\end{equation}
where $S_{{\rm{BZ}}}$ is the BZ area. Fig. \ref{fig3} shows the relation between the quantities $\tilde \lambda _C /t$ and $\Delta /t$ determined by solving Eq. (\ref{eq41}). Both quantities have a nearly linear dependence on each other. The coupling parameter $\Delta$ corresponding to $\tilde \lambda _C /t = 1.307$ is $\Delta  = 0.099t$, which yields the bulk gap of $E_g  = 0.156t$. If we set the hopping parameter to the value of graphene ($t = 2.8{\rm{eV}}$), then the system will have the bulk gap that is one order of magnitude larger than the ones of typical 2D topological insulators\cite{ref38}.

\begin{figure}[h!]
	\begin{center}
		\includegraphics[width=13.0cm]{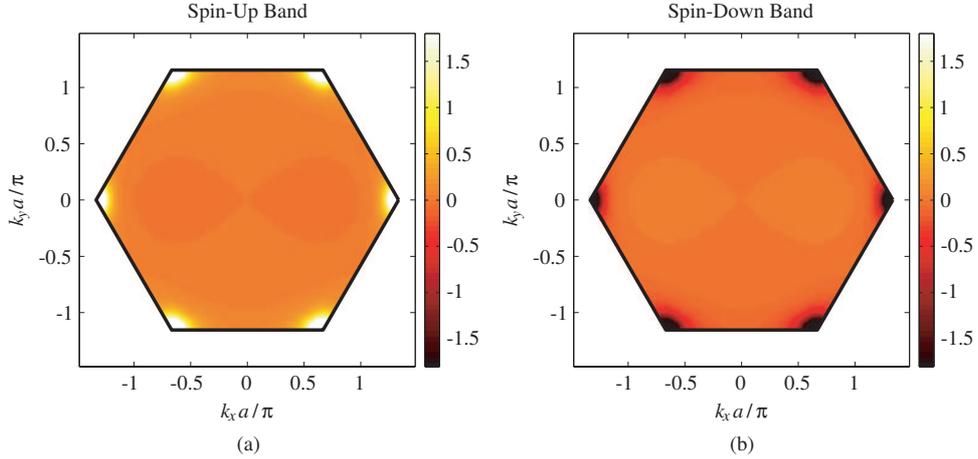}
	\end{center}
	\caption{(Color online) Distributions of the dimensionless Berry curvatures $\tilde \Omega _\sigma  ({\bf{k}}) = \frac{1}{{a^2 }}\Omega _\sigma  ({\bf{k}})$ for (a) the spin-up band and (b) the spin-down band. There are strong peaks at the ${\rm{K}}$ and ${\rm{K'}}$ points.}
	\label{fig4}
\end{figure}

The topology of the system can be characterized by the Chern number. We can define the separate Chern number for the half-filled spin-$\sigma$ electrons by
\begin{equation}\label{eq42}
n_{{\rm{C}},\sigma }  \equiv \frac{1}{{2\pi }}\int_{{\rm{BZ}}} {d^2 k}  {\kern 2pt} \Omega _\sigma  ({\bf{k}}).
\end{equation}
Here the Berry curvature $\Omega _\sigma$ is defined as
\begin{equation}\label{eq43}
\Omega _\sigma  ({\bf{k}}) \equiv i\left( {\left\langle {{\frac{{\partial \psi _\sigma ^ -  ({\bf{k}})}}{{\partial k_x }}}}
	\mathrel{\left | {\vphantom {{\frac{{\partial \psi _\sigma ^ -  ({\bf{k}})}}{{\partial k_x }}} {\frac{{\partial \psi _\sigma ^ -  ({\bf{k}})}}{{\partial k_y }}}}}
		\right. \kern-\nulldelimiterspace}
	{{\frac{{\partial \psi _\sigma ^ -  ({\bf{k}})}}{{\partial k_y }}}} \right\rangle  - \left\langle {{\frac{{\partial \psi _\sigma ^ -  ({\bf{k}})}}{{\partial k_y }}}}
	\mathrel{\left | {\vphantom {{\frac{{\partial \psi _\sigma ^ -  ({\bf{k}})}}{{\partial k_y }}} {\frac{{\partial \psi _\sigma ^ -  ({\bf{k}})}}{{\partial k_x }}}}}
		\right. \kern-\nulldelimiterspace}
	{{\frac{{\partial \psi _\sigma ^ -  ({\bf{k}})}}{{\partial k_x }}}} \right\rangle } \right).
\end{equation}
For the half-filled two-band system with the Hamiltonian in Eq. (\ref{eq39}), the Chern number can be calculated using simple formula\cite{ref39},
\begin{equation}\label{eq44}
n_{{\rm{C}},\sigma }  = \frac{1}{{4\pi }}\int_{{\rm{BZ}}} {d^2 k} {\kern 2pt} {\bf{\hat d}}_\sigma  ({\bf{k}}) \cdot \left( {\frac{{\partial {\bf{\hat d}}_\sigma  ({\bf{k}})}}{{\partial k_x }} \times \frac{{\partial {\bf{\hat d}}_\sigma  ({\bf{k}})}}{{\partial k_y }}} \right),
\end{equation}
with the unit vector ${\bf{\hat d}}_\sigma  ({\bf{k}}) = {\bf{d}}_\sigma  ({\bf{k}})/|{\bf{d}}_\sigma  ({\bf{k}})|$. We simply call the two-band system of the spin-up (spin-down) electrons as a \emph{spin-up (spin-down) band}. The distribution of the Berry curvature has been considered. Fig. \ref{fig4} shows the results for the spin-up and spin-down bands. The peaks of the curvatures are located at the ${\rm{K}}$ and ${\rm{K'}}$ points where the modulus $|{\bf{d}}_\sigma  ({\bf{k}})|$ has its minimum.

\begin{figure}[h!]
	\begin{center}
		\includegraphics[width=13.0cm]{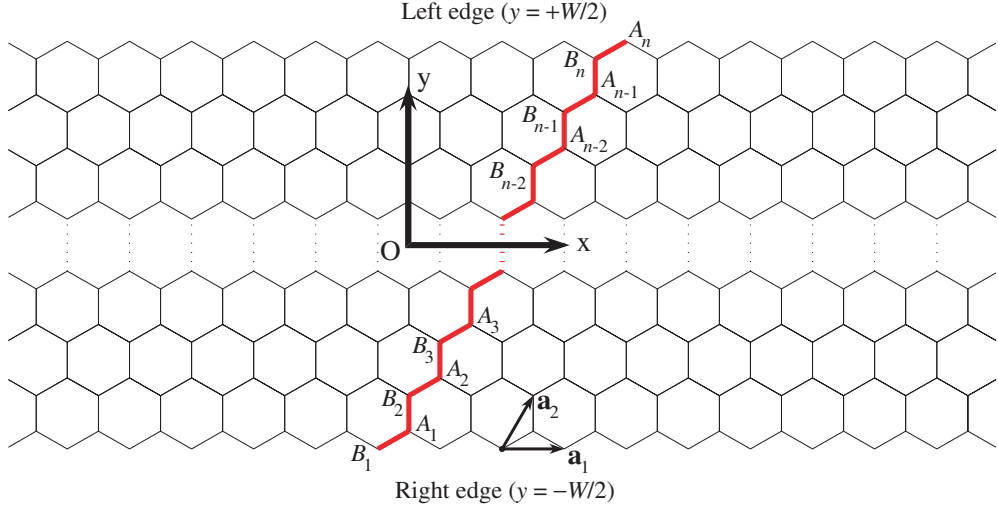}
	\end{center}
	\caption{(Color online) Strip geometry of the honeycomb lattice with zigzag edges along the $x$-axis. Here $W$ is the width of the geometry and $n$ is the number of the unit cells in the $y$-direction. The thick red line indicates a 1D unit cell of the system containing $2n$ atoms.}
	\label{fig5}
\end{figure}

The numerically calculated Chern numbers for the spin-up and spin-down bands are $n_{{\rm{C}},\sigma  =  \pm 1}  =  \pm 0.99995$, indicating the exact value of $n_{{\rm{C}},\sigma  =  \pm 1}  =  \pm 1$. The nontrivial Chern numbers are necessarily linked with the emergence of the edge states. In order to determine the edge states, we calculate the one-dimensional (1D) band structure for the strip geometry of the honeycomb lattice shown in Fig. \ref{fig5}.
The 1D band structures can be found from the real-space Hamiltonian which is obtained by a Fourier transformation of the 2D Hamiltonian (\ref{eq39}). Inserting $c_{{\bf{k}},o,\sigma }  = \frac{1}{{\sqrt N }}\sum\limits_i {c_{io\sigma } e^{ - i{\bf{k}} \cdot {\bf{R}}_i } }$ into Eq. (\ref{eq39}), we have the Hamiltonian in real space,
\begin{equation}\label{eq45}
\begin{split}	
H_\sigma ^{{\rm{MF}}}  = & - t\sum\limits_{\left\langle {iA,jB} \right\rangle } {(c_{iA\sigma }^ \dagger  c_{jB\sigma }  + {\rm{H}}{\rm{.c}}{\rm{.}})}  \\
& - \Delta \sigma \sum\limits_i {\sum\limits_{m = 2,4,6} {[ig_1 (c_{iA\sigma }^ \dagger  c_{{\bf{R}}_i  - {\bf{R}}_m ,A,\sigma }  - c_{iB\sigma }^ \dagger  c_{{\bf{R}}_i  - {\bf{R}}_m ,B,\sigma } ) + {\rm{H}}{\rm{.c}}{\rm{.]}}} } \\
& + \Delta \sigma \sum\limits_i {\sum\limits_{m = 14,16,18} {[ig_2 (c_{iA\sigma }^ \dagger  c_{{\bf{R}}_i  - {\bf{R}}_m ,A,\sigma }  - c_{iB\sigma }^ \dagger  c_{{\bf{R}}_i  - {\bf{R}}_m ,B,\sigma } ) + {\rm{H}}{\rm{.c}}{\rm{.]}}} }.
\end{split}	
\end{equation}
At $g_2  = 0$, it becomes the Kane-Mele model\cite{ref01, ref02} with the mirror and inversion symmetries.

\begin{figure}[h!]
	\begin{center}
		\includegraphics[width=13.0cm]{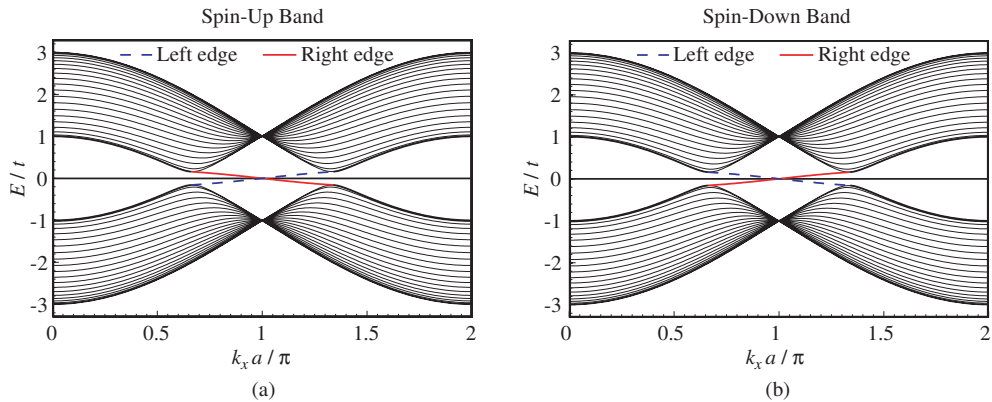}
	\end{center}
	\caption{(Color online) 1D band structures for (a) spin-up electrons and (b) spin-down electrons in the strip geometry with zigzag edges. The spin-up and spin-down electrons move in opposite directions at the same edge, i.e., they have opposite chiralities.}
	\label{fig6}
\end{figure}

And then, we need to Fourier transform the Hamiltonian (\ref{eq45}) only in the $x$-axis to get the 1D Hamiltonian matrix $[H_\sigma ^{{\rm{MF}}} (k_x )]_{oi,o'j}$. Here the sublattice indices $o$ and $o'$ take two values, $A$ or $B$, while the indices $i$ and $j$ run from 1 to $n$ (the number of the unit cells in the $y$-direction as shown in Fig. \ref{fig5}). By numerical diagonalization of this $2n \times 2n$ matrix we obtain the 1D band structure. The results for the spin-up and spin-down bands are demonstrated in Fig. \ref{fig6}.

For the range of the wave number between $k_x a = \frac{{2\pi }}{3}$ and $k_x a = \frac{{4\pi }}{3}$, there exist four groups of the sub-gap states that are localized near both edges. They are dubbed the \emph{edge modes}. As one can see from the dispersion relation of the edge modes, the electrons at the same edge, but with opposite spin polarities, propagate in opposite directions. So the edge states of the spin-up and spin-down electrons have opposite chiralities, which is called the \emph{helical edge states}. Since the total Chern number is zero ($n_{\rm{C}}  = n_{{\rm{C}},1}  + n_{{\rm{C}}, - 1}  = 0$), the system does not support the quantum Hall effect. However, a dissipationless spin current can exist on the edges of the system, making it possible that the system exhibits the quantum spin Hall effect.

Lastly we comment on the MF theory. The MF approach has a well-known shortcoming, i.e., it neglects completely the influence of any fluctuations of the order parameters. This neglecting can lead to the overestimation of the ordering tendencies. As is well known from statistical mechanics, an order parameter associated with spontaneously broken continuous symmetry can excite low energy modes called the Goldstone modes, and the fluctuations of these modes can destroy corresponding long-range order. For example, due to the Mermin-Wagner theorem\cite{ref41}, the spontaneous breaking of the spin-SU(2) symmetry is impossible, in a strict sense, at finite temperature in two-dimensional systems. However, it has been argued in a previous work\cite{ref42} that, in the system with finite macroscopic size, the long-range order in spin channel would survive at the temperature, which is considerably lower than the MF-predicted critical temperature, but is still finite. Thus, according to the argument of this work, the Mermin-Wagner theorem is not practically applicable for the spontaneous breaking of the continuous spin-rotation symmetry in the spin-ordered states of two-dimensional systems, despite its formal correctness. In this paper, we addressed only the case of the zero-temperature limit. A similar effect of the fluctuations can affect the results here, e.g., the gap size and the form-factor structure of the order parameter, but the investigation of this problem is beyond the scope of the present work.

\section{Conclusion}\label{sec5}

In the present work, we have used the TUFRG to detect and analyze the QSH state of the half-filled honeycomb lattice. It has addressed two problems.

First, we have revisited the half-filled honeycomb lattice with enhanced exchange coupling and presented the TUFRG phase diagram with higher resolution in the parameter space than that in previous work\cite{ref20}. It depicts more precisely the boundary of the QSH phase and can provide useful hint for the search for the topological Mott insulator.

Second, we have proposed a scheme for linking the TUFRG with MF theory. In our scheme, only the irreducible part of the effective interaction from TUFRG is employed as an input interaction for the MF analysis. It has been applied to analyze in detail the QSH phase at $V_2  = t$ and $J = 2t$. The phase has the bulk gap of $E_g  = 0.156t$. If we assume the parameter $t$ to be the value of graphen, then the gap would be one order of magnitude larger than those of typical 2D topological insulators\cite{ref38}, which is of crucial importance for its application.

Although the present work is limited to the case where the effective interaction has only one singular mode in a single channel, but we think, it would not be difficult to extend our argument to the more complicated case like the coexistence phase. This task would be our future work.

\section*{Acknowledgments}
We thank Chol-Jun Kang for useful discussions.

\section*{References}
\bibliography{SJO_Manuscript}

\end{document}